# Aftershocks and fluctuating diffusivity

Sumiyoshi Abe [1,2,3,*], Norikazu Suzuki [4], Dmitrii A. Tayurskii [2]

[1] Department of Physics, College of Information Science and Engineering,
   Huaqiao University, Xiamen 361021, China

[2] Institute of Physics, Kazan Federal University, Kazan 420008, Russia

[3] Department of Natural and Mathematical Sciences, Turin Polytechnic University
   in Tashkent, Tashkent 100095, Uzbekistan

[4] College of Science and Technology, Nihon University, Chiba 274-8501, Japan

[*] Correspondence: suabe@sf6.so-net.ne.jp

**Abstract:** The Omori-Utsu law shows the temporal power-law-like decrease of the frequency of earthquake aftershocks and, interestingly, is found in a variety of complex systems/phenomena exhibiting catastrophes. Now, it may be interpreted as a characteristic response of such systems to large events. Here, hierarchical dynamics with the fast and slow degrees of freedom is studied on the basis of the Fokker-Planck theory for the load-state distribution to formulate the law as a relaxation process, in which diffusion coefficient in the space of the load state is treated as a fluctuating slow variable. The evolution equation reduced from the full Fokker-Planck equation and its Green's function are analyzed for the subdynamics governing the load state as the fast degree of freedom. It is shown that the subsystem has the temporal translational invariance in the logarithmic time, not in the conventional time, and consequently the aging phenomenon appears.

## 1. Introduction

Given a system, it is essential for understanding its properties out of equilibrium to look at how the system responds to a sudden disturbance and returns to a typical state such as the equilibrium one. Therefore, the relaxation phenomena are of major interest for a stable system. Most celebrated may be the exponential relaxation, to the type of which the mechanical stress relaxation of Maxwell [1] and the Drude-Lorentz-Debye dielectric(-paramagnetic) relaxation [2,3] belong. However, it is also known that there exist systems of another type characterized by the stretched exponential function exhibiting temporal decays of disturbances slower than the exponential type. Such an exotic type has been identified already in the 19th century [4]. It is termed the Kohlrausch-Williams-Watts relaxation and is widely observed in disordered systems such as glass-forming liquids (Reference [5] for a review) and polymers (see Reference [6], for example). Furthermore, it has been found [7,8] that there are soft matters obeying even slower power-law stress relaxation.

The power-law relaxation phenomenon has actually been long known in seismology. Omori [9] has discovered in the 19th century that the temporal decrease of aftershocks following a large earthquake obeys a power law. Later, this empirical law has been modified and made more precise by Utsu [10]. Accordingly, today it is commonly referred to as the Omori-Utsu law for earthquake aftershocks. It is as follows. Suppose that a main shock has occurred at $t=0$. Then, the law states that the number of aftershocks occurring during the later time interval between $t$ and $t+\Delta t$, $\Delta N(t) = N(t+\Delta t) - N(t)$, is $\Delta N(t) \sim (c+t)^{-p} \Delta t$ with $p$ and $c$ being positive constants.

In the continuous approximation, it is written in the form of differential as

$$\frac{d N(t)}{d t} = n_0 \, \phi(t), \tag{1}$$

where $n_0$ is a positive constant, and $\phi(t)$ is the function describing the power-law relaxation

$$\phi(t) = \frac{1}{(1 + t/\tau)^p} \tag{2}$$

with $\tau = c$. The value of the exponent $p$ is generally considered to be close to unity but actually ranges between about 0.5 and 1.5, depending on datasets.

An interesting point is that the Omori-Utsu law has its analogs outside seismology. A couple of examples have been observed in connection with the World Wide Web. It has been reported [11] that the download rate of an article uploaded online obeys the Omori-Utsu-like law. It has also been shown [12] that the pattern of the Internet traffic after a heavily congested state also obeys it. Furthermore, analogous phenomena have been discovered in the behaviors of the financial markets. In Reference [13], it has been found that the dynamics of the stock market after a large crash is well described by the Omori-Utsu-like law. In addition, it has also been demonstrated to be the case concerning the crash of the currency exchange rate [14]. These facts suggest that the Omori-Utsu law may serve as a key for understanding complex systems with catastrophes, in general.

In this paper, we develop a discussion about the Omori-Utsu law for occurrence of earthquake aftershocks based on physical kinetics. In particular, a generalization of the Fokker-Planck theory, in which the diffusion coefficient is not a constant but a fluctuating variable that may describe the effects originating from heterogeneity of the crust and a complex landscape of the stress distribution at faults, and see how such an theory can explain the features of the temporal pattern of aftershocks. For this purpose, "a modified version" of the approach recently proposed in Reference [15] is employed for representing the dynamical hierarchy. It is shown that the relaxation function in Equation (2) can be obtained for the subsystem defined through elimination of the fluctuating diffusivity. This, in turn, gives information on the dynamics underlying the aftershock phenomenon. Then, we study the reduction of the generalized Fokker-Planck equation and analyze associated Green's function characterizing the transition of the load state. It is found that Green's function has the temporal translational invariance not in the conventional time but in the logarithmic time. As a result, the system exhibits the phenomenon of aging, that is, the system has its own clock. This is in conformity with the discovery in real seismicity reported in References [16,17]. On the other hand, time evolution is however still Markovian since Green's function obeys the Chapman-Kolmogorov equation, in contrast to the non-Markovian nature of aftershocks [18] (see also Reference [17]). Thus, the present theory can partially describe the properties of aftershocks.

## 2. Omori-Utsu Law and Hierarchical Fokker-Planck Equation with Fluctuating Diffusivity

The concept of Brownian motion has been discussed in the context of seismology in References [19,20] (see also other works cited therein). There, recurrence of earthquakes has been modeled as a problem of passage times [21]. The basic random variable considered there, which experiences random perturbations, is termed "load state" [20]. In the context of statistical mechanics, it may be thought of as the mean field of stresses distributed over faults in a geographical region under consideration. We shall denote its realization by $\xi$. Thus, *$\xi$ should not be confused with a spatial coordinate*.

Our point is as follows. We hypothesize that the load state in the spatial regime of aftershocks undergoes the heterogeneity described by the fluctuating diffusivity, implying that the diffusion coefficient *D* itself is regarded as the realization of a random variable, not a fixed constant. On the other hand, such exotic diffusion with fluctuating diffusivities has been observed in laboratories [22,23], where the heterogeneities of the systems are essential. To theoretically describe it, a kinetic approach has been developed in Reference [15] on the basis of the Fokker-Planck theory endowed with the hierarchical dynamics characterized by largely separated time scales. In view of that approach, the load-state variable $\xi$ is the fast degree of freedom, whereas the diffusivity variable *D* should be the slow degree of freedom.

Thus, we consider a 2-tuple of the dynamical random variables $\mathbf{X} = (X_1, X_2)^T$, which obeys a general stochastic differential equation [24]: $d\mathbf{X} = \tilde{\mathbf{K}}\, dt + \tilde{G}\, d\mathbf{W}$. $\tilde{\mathbf{K}}$ is a drift term and $\tilde{G}$ is a $2 \times 2$ matrix. Both of them depend on $X_1$, $X_2$ and time *t*. $d\mathbf{W} = (dW_1, dW_2)^T$ is assumed to be the Wiener noise satisfying Itô's rule:

$dW_i dW_j = \delta_{ij} dt$. This implies that $dW_1$ and $dW_2$ are mutually independent, but a possible correlation between them may effectively be taken into account by a nondiagonal $\tilde{G}$. Now, $X_1$ is taken to be the random variable of the load state whose realization is $\xi$ and $X_2$ is the logarithm of the nonnegative random variable of diffusivity with its realization $D$. Since the diffusivity variable is dimensioned, its logarithm contains a constant eliminating the dimensionality. However, without loss of generality, such a constant can be set equal to unity. Therefore, the realization $\mathbf{x}$ of $\mathbf{X}$ is denoted by $\mathbf{x} = (x_1, x_2)^T = (\xi, \ln D)^T$.

Our interest is in determining $\tilde{\mathbf{K}}$ and $\tilde{G}$ that lead to the Omori-Utsu law since these quantities carry information on the dynamics underlying the law. In this respect, the Fokker-Planck theory may give a clue. In what follows, we examine this point based on a modification of the method presented in Reference [15].

Let $\tilde{P}(x_1, x_2, t)$ be the probability distribution normalized in the whole plane: $-\infty < x_1 < \infty$, $-\infty < x_2 < \infty$. Physically, this infinite support can be approximated to be a finite one for the well-localized system/phenomenon. See the later discussion about Equation (31). This distribution obeys the Fokker-Planck equation [24] associated with the above-mentioned stochastic differential equation in the following general form:

$$\frac{\partial \tilde{P}}{\partial t} = -\sum_{i=1,2} \frac{\partial}{\partial x_i}\left(\tilde{K}_i \tilde{P}\right) + \sum_{i,j=1,2} \frac{\partial^2}{\partial x_i \partial x_j}\left(\tilde{\sigma}_{ij} \tilde{P}\right), \qquad (3)$$

where $\tilde{\sigma} = (\tilde{\sigma}_{ij})$ is a positive matrix (i.e. being symmetric and having only positive eigenvalues) given by $\tilde{\sigma} = (1/2)\tilde{G}\tilde{G}^T$, and both $\tilde{K}_i$ and $\tilde{\sigma}_{ij}$ are the functions of $x_1$, $x_2$ and $t$. Mathematically, the drift term has the dependency on calculi since the underlying stochastic process mentioned above is multiplicative. Here, Itô calculus has been employed.

To be explicit, it is convenient to directly employ the pair $(\xi, D)$. Accordingly, $\partial/\partial x_1 = \partial/\partial \xi$, $\partial/\partial x_2 = D\partial/\partial D$ and $P(\xi, D, t) = (1/D)\tilde{P}(\xi, \ln D, t)$ with the prefactor $1/D$ being the Jacobian. $P(\xi, D, t)$ is normalized in the upper-half plane since $D$ is nonnegative. Then, Equation (1) is rewritten as

$$\frac{\partial P}{\partial t} = -\frac{\partial}{\partial \xi}(K_1 P) - D\frac{\partial}{\partial D}(K_2 P) \\ + \frac{\partial^2}{\partial \xi^2}(\sigma_{11} P) + 2\frac{\partial^2}{\partial \xi \partial D}(\sigma_{12} P) + \frac{\partial^2}{\partial D^2}(\sigma_{22} P), \qquad (4)$$

which is "a modification" of Equation (13) in Reference [15]. In this equation, $K$'s and $\sigma$'s may depend on $\xi$ and $D$ as well as $t$ and are related to $\tilde{K}$'s and $\tilde{\sigma}$'s in Equation (3) as follows: $K_1 = \tilde{K}_1$, $K_2 = D(\tilde{K}_2 + \tilde{\sigma}_{22})$, $\sigma_{11} = \tilde{\sigma}_{11}$, $\sigma_{12} = D\tilde{\sigma}_{12}$, $\sigma_{22} = D^2\tilde{\sigma}_{22}$. As mentioned, $\xi$ and $D$ are the fast and slow degrees of freedom, respectively. To implement such a hierarchy, the method analogous to the Born-Oppenheimer approximation may be applied [15]. That is, *the fast degree of freedom is strongly influenced by the slow degree of freedom, whereas the slow degree of freedom is not affected by the fast degree of freedom*. This leads to

$$K_2 = K_2(D), \qquad \sigma_{22} = \sigma_{22}(D), \tag{5}$$

provided that time dependence in these quantities are ignored since they are relevant to the slow degree of freedom. In addition, the joint probability distribution should be factorized as follows:

$$P(\xi, D, t) = p(\xi, t \mid D) \Pi(D), \tag{6}$$

where $p(\xi, t \mid D)$ is the conditional probability distribution given the value of $D$ and $\Pi(D)$ is the marginal probability distribution of $D$, time dependence of which is ignored as in Equation (5). Then, Equation (4) becomes

$$\begin{aligned}
\Pi(D)\frac{\partial p(\xi, t \mid D)}{\partial t} = &-\Pi(D)\frac{\partial}{\partial \xi}\left[K_1(\xi, D, t) p(\xi, t \mid D)\right] \\
&-\frac{\partial}{\partial D}\left[K_2(D) p(\xi, t \mid D) \Pi(D)\right] \\
&+\Pi(D)\frac{\partial^2}{\partial \xi^2}\left[\sigma_{11}(\xi, D, t) p(\xi, t \mid D)\right] \\
&+2\frac{\partial}{\partial D}\left\{\Pi(D)\frac{\partial}{\partial \xi}\left[\sigma_{12}(\xi, D, t) p(\xi, t \mid D)\right]\right\} \\
&+\frac{\partial^2}{\partial D^2}\left[\sigma_{22}(D) p(\xi, t \mid D) \Pi(D)\right].
\end{aligned} \tag{7}$$

As shown in Reference [15] with the present modification, this equation can be separated as follows:

$$\frac{\partial p(\xi,t|D)}{\partial t} = -\frac{\partial}{\partial \xi}\left[K_1(\xi,D,t)p(\xi,t|D)\right]$$

$$+\frac{\partial^2}{\partial \xi^2}\left[\sigma_{11}(\xi,D,t)p(\xi,t|D)\right] \qquad (8)$$

for the fast degree of freedom and

$$-\frac{\partial}{\partial D}\left[K_2(D)p(\xi,t|D)\Pi(D)\right]+2\frac{\partial}{\partial D}\left\{\Pi(D)\frac{\partial}{\partial \xi}\left[\sigma_{12}(\xi,D,t)p(\xi,t|D)\right]\right\}$$

$$+\frac{\partial^2}{\partial D^2}\left[\sigma_{22}(D)p(\xi,t|D)\Pi(D)\right]=0 \qquad (9)$$

for the rest. Equation (9) is immediately integrated to be

$$-K_2(D)p(\xi,t|D)\Pi(D)+2\Pi(D)\frac{\partial}{\partial \xi}\left[\sigma_{12}(\xi,D,t)p(\xi,t|D)\right]$$

$$+\frac{d\left[\sigma_{22}(D)\Pi(D)\right]}{dD}p(\xi,t|D)+\sigma_{22}(D)\Pi(D)\frac{\partial p(\xi,t|D)}{\partial D}=f(\xi,t). \qquad (10)$$

Here, $f(\xi,t)$ is a certain function. This equation is further separated as follows:

$$-K_2(D)\Pi(D)+\frac{d\left[\sigma_{22}(D)\Pi(D)\right]}{dD}=0 \qquad (11)$$

for the slow degree of freedom and

$$2\Pi(D)\frac{\partial}{\partial \xi}\left[\sigma_{12}(\xi,D,t)p(\xi,t|D)\right]+\sigma_{22}(D)\Pi(D)\frac{\partial p(\xi,t|D)}{\partial D}=f(\xi,t) \qquad (12)$$

for the coupling between the fast and slow degrees of freedom. Actually, it is possible to set

$$f(\xi, t) = 0. \tag{13}$$

This is because the integration of Equation (12) over $-\infty < \xi < \infty$ leads to the fact that such an integration of $f(\xi,t)$ vanishes, provided that $\sigma_{12}(\xi, D, t) p(\xi, t | D) \to 0$ ($\xi \to \pm\infty$) is assumed. [There is still a possibility that $f(\xi,t)$ is an integrable odd function of $\xi$, but such a case turns out to be ruled out.] Thus, the equation for the coupling is given by

$$2 \frac{\partial}{\partial \xi} \left[ \sigma_{12}(\xi, D, t) p(\xi, t | D) \right] + \sigma_{22}(D) \frac{\partial p(\xi, t | D)}{\partial D} = 0. \tag{14}$$

As required, the slow degree of freedom does not depend on the fast degree of freedom in Equation (11). Equations (8), (11) and (14) characterize the hierarchical dynamics. Let us analyze these for the Omori-Utsu law.

Firstly, we discuss Equation (8) for the fast degree of freedom. Since the external loading on the region under consideration such as the tectonic one [19] is negligible, we impose the condition on the drift term in Equation (8) that it is actually absent:

$$K_1 = 0. \tag{15}$$

Therefore, we have

$$\frac{\partial p(\xi, t | D)}{\partial t} = \frac{\partial^2}{\partial \xi^2} \left[ \sigma_{11}(\xi, D, t) p(\xi, t | D) \right]. \tag{16}$$

Following the Brownian model of seismicity proposed in References [19,20], we take the Gaussian solution to Equation (16):

$$p(\xi,t|D) = \frac{1}{\sqrt{4\pi Dt}} \exp\left(-\frac{\xi^2}{4Dt}\right), \qquad (17)$$

which corresponds to the initial condition $p(\xi,0|D) = \delta(\xi)$, representing a main shock at $t=0$. Although a more general initial condition may actually be employed, here we consider this simplest one. Substituting Equation (17) into Equation (16), we have

$$\sigma_{11} = D, \qquad (18)$$

as expected.

Let us see how Equation (17) can give rise to the Omori-Utsu law. Recall that the relaxation function in a symmetric random walk model is given by the characteristic function of the distribution [25-27]. In the present case, it is the characteristic function of the marginal distribution of $\xi$ obtained by the integration of the joint distribution $P(\xi,D,t)$ over $D$. That is,

$$\phi_k(t) = \int_0^\infty dD\, \chi(k,t|D)\Pi(D). \qquad (19)$$

In this equation, $\chi(k,t|D)$ stands for the characteristic function of the conditional distribution

$$\chi(k,t|D) = \int_{-\infty}^{\infty} d\xi \, e^{ik\xi} \, p(\xi,t|D), \tag{20}$$

which is calculated for Equation (17) to be

$$\chi(k,t|D) = \exp(-Dk^2 t). \tag{21}$$

Let us examine the case when the fluctuating diffusivity obeys the gamma distribution

$$\Pi(D) = \frac{1}{\Gamma(p)} \frac{D^{p-1}}{D_0^p} \exp\left(-\frac{D}{D_0}\right) \quad (p > 0), \tag{22}$$

where $\Gamma(p)$ is the gamma function and $D_0$ is a constant given by $D_0 = \langle D \rangle / p$ with $\langle Q \rangle \equiv \int_0^\infty dD \int_{-\infty}^\infty d\xi \, Q P(\xi,D,t)$, Equations (6), (17) and (22). Then, from Equations (19)-(21), we obtain the relaxation function for the Omori-Utsu law

$$\phi_k(t) = \frac{1}{(1 + t/\tau_k)^p}, \tag{23}$$

where $\tau_k$ is given by

$$\tau_k = \frac{1}{D_0 k^2}. \tag{24}$$

Thus, Equation (2) is in fact obtained for each mode $k$. This is our first result.

Secondly, let us analyze Equation (11). Using Equation (22), that equation leads to

the following relation:

$$K_2(D) = \frac{d\sigma_{22}(D)}{dD} + \left(\frac{p-1}{D} - \frac{1}{D_0}\right)\sigma_{22}(D). \tag{25}$$

$\sigma_{22}$ will be determined later.

Thirdly, substituting Equation (17) into Equation (14), we have the following equation for the coupling between the fast and slow degrees of freedom:

$$2\frac{\partial \sigma_{12}(\xi,D,t)}{\partial \xi} - \frac{\xi}{Dt}\sigma_{12}(\xi,D,t) + \left(\frac{\xi^2}{4D^2 t} - \frac{1}{2D}\right)\sigma_{22}(D) = 0. \tag{26}$$

This equation has the solution

$$\sigma_{12}(\xi, D) = \frac{\xi}{4D}\sigma_{22}(D), \tag{27}$$

which shows that the coupling is fixed in time.

Finally, as in Reference [15], we determine $\sigma_{22}$ by the positivity of $\sigma$. Using Equations (5), (18) and (27), this matrix is written as

$$\sigma = \begin{pmatrix} D & [\xi/(4D)]\sigma_{22}(D) \\ [\xi/(4D)]\sigma_{22}(D) & \sigma_{22}(D) \end{pmatrix}. \tag{28}$$

Since the positivity implies that the eigenvalues of this $2\times 2$ matrix is positive, it follows that

$$\mathrm{tr}\,\sigma = D + \sigma_{22}(D) > 0, \tag{29}$$

$$\det \sigma = \sigma_{22}(D)\left[D - \frac{\xi^2}{16D^2}\sigma_{22}(D)\right] > 0. \tag{30}$$

Equation (29) is natural, whereas Equation (30) needs some considerations. Clearly, in order for the quantity inside the square brackets in Equation (30) to be positive, $\xi$ cannot arbitrarily be large. This, in turn, imposes a constraint on the time scale for the validity of the present theory [15]. Let such a time scale be denoted by $T$. The diffusion property suggests that $\xi^2 \sim \langle \xi^2 \rangle = 2pD_0 T$, where $\langle \xi^2 \rangle = \int_{-\infty}^{\infty} d\xi \int_{0}^{\infty} dD\, \xi^2 P(\xi, D, T)$ is the variance of $\xi$ in terms of the distribution in Equation (6) with Equations (17) and (22) and $\langle \xi \rangle = 0$ (see the last paragraph in this section). This may indicate that the corresponding scale $S$ is a constant satisfying

$$\sqrt{2pD_0 T} \ll S. \tag{31}$$

In other words, the value of $\xi$ should be well localized in this way. Thus, Equation (30) may hold up to $S$ (or correspondingly $T$). In terms of such a scale, we have

$$\sigma_{22}(D) = \frac{16D^3}{S^2} \tag{32}$$

as the solution. Substituting this equation into Equations (25) and (27), we obtain

$$K_2(D) = (p+2)\frac{16D^2}{S^2} - \frac{16D^3}{D_0 S^2}, \tag{33}$$

$$\sigma_{12}(\xi, D) = \frac{4\xi D^2}{S^2}, \tag{34}$$

respectively.

Consequently, we find that the present theory based on the Fokker-Planck equation with fluctuating diffusivity describes the Omori-Utsu law if $K$'s and $\sigma$'s are given by Equations (15), (18) and (32)-(34).

Closing this section, we present the explicit form of the marginal distribution of $\xi$, which is denoted here by $\hat{p}(\xi, t)$. As seen in Equations (19) and (20), it is given by the inverse Fourier transformation of Equation (23):

$$\hat{p}(\xi, t) = \frac{1}{2\pi} \int_{-\infty}^{\infty} dk\, e^{-ik\xi} \phi_k(t). \tag{35}$$

Then, using the formula (9.6.25) in Reference [28], we have

$$\hat{p}(\xi, t) = \frac{1}{2^{p-1/2} \sqrt{\pi}\, \Gamma(p) \sqrt{D_0 t}} \left( \frac{|\xi|}{\sqrt{D_0 t}} \right)^{p-1/2} K_{p-1/2}\left( \frac{|\xi|}{\sqrt{D_0 t}} \right), \tag{36}$$

where $K_\nu(z)$ is the modified Bessel function. The $\xi$-dependence appears only in the combined form $|\xi|/\sqrt{D_0 t}$, implying the normal diffusion: the variance of $\xi$ linearly grows in time $t$. Therefore, the present model offers an example of the non-Gaussian normal diffusion, the phenomenon of which is currently attracting much attention [22,23]. It is however noted that this diffusion process takes place in *the space of the*

*load state*, not in the conventional space.

## 3. Subdynamics, Logarithmic Time and Aging

Now, we address ourselves to studying the subdynamics obtained by reduction of the fluctuating diffusivity.

Let us derive the evolution equation for the marginal distribution $\hat{p}(\xi,t)$. It is should be noted that such an equation is necessarily specific to the initial condition. Equation (17) satisfies

$$\frac{\partial p(\xi,t|D)}{\partial t} = -\frac{1}{2t}\left(\xi\frac{\partial}{\partial \xi}+1\right)p(\xi,t|D). \tag{37}$$

With this form, multiplying the both side by $\Pi(D)$ and integrating over $D$, we have the following equation for the marginal distribution:

$$\frac{\partial \hat{p}(\xi,t)}{\partial t} = -\frac{1}{2t}\left(\xi\frac{\partial}{\partial \xi}+1\right)\hat{p}(\xi,t), \tag{38}$$

which describes how the subsystem evolves in time.

Here, we wish to make a comment on the fact that the operator in Equation (37) does not have explicit dependence on *D* and accordingly the marginal distribution satisfies a closed form as in Equation (38). In fact, upon deriving that equation, we do not have to assume the explicit form of $\Pi(D)$ in Equation (22). Actually, this feature has its origin in the scaling property of the conditional distribution in Equation (17):

$$p(\xi,t\,|\,D) = \frac{1}{t^{1/2}}\,\bar{p}(\xi/t^{1/2}\,|\,D), \tag{39}$$

where $\bar{p}(x\,|\,D)$ is the Gaussian scaling function $\bar{p}(x\,|\,D) = (4\pi D)^{-1/2}\exp\left[-x^2/(4D)\right]$. In this context, the scale invariance of Equation (38) should be noted: it does not change its form under the individual rescaling transformations of $\xi$ and $t$. In Section 4, a further discussion will be made about the relation between the scaling property and derivability of an evolution equation for a marginal distribution in a closed form.

To see the property of the subdynamics, let us analyze Green's function $G(\xi,t:\xi',t')$ associated with Equation (38), which is the solution of the equation

$$\left[\frac{\partial}{\partial t} + \frac{1}{2t}\left(\xi\frac{\partial}{\partial \xi} + 1\right)\right]G(\xi,t:\xi',t') = \delta(\xi-\xi')\delta(t-t'), \tag{40}$$

satisfying the condition

$$\lim_{t\to t'+0} G(\xi,t:\xi',t') = \delta(\xi-\xi'). \tag{41}$$

The explicit form of the solution is found to be given by

$$G(\xi,t:\xi',t') = \frac{1}{t^{1/2}}\delta\left(\xi/t^{1/2} - \xi'/t'^{1/2}\right)\theta(t-t'), \tag{42}$$

where $\theta(s)$ is the Heaviside step function: $\theta(s) = 0\,(s<0)$, $1/2\,(s=0)$, $1\,(s>0)$.

From Equation (42), we obtain three important results, which are as follows.

Firstly, *the transition from one value of the load state to another is deterministic*

because of the delta-function nature of Green's function. This is due to the fact that Equation (40) does not depend on $D_0$: no remnants of the diffusivity are contained. The functional form in Equation (36) is kept unchanged under time evolution (as discussed in Section 4, this comes from the scaling property of the conditional distribution that still depends on *D*).

Secondly, Green's function clearly satisfies the Chapman-Kolmogorov equation [24]

$$G(\xi,t:\xi',t') = \int_{-\infty}^{\infty} d\xi'' G(\xi,t:\xi'',t'') G(\xi'',t'':\xi',t') \quad (t > t'' > t'). \quad (43)$$

Therefore, the evolution process is Markovian.

Thirdly, the evolution is not stationary since the dependence of Green's function on $t$ and $t'$ cannot be expressed in terms only of the difference $t - t'$ and so the temporal translational invariance is violated. This is actually clear since the operator in Equation (40) has explicit time dependence. However, it is of significance to rewrite Equation (42) in the following form:

$$G(\xi,t:\xi',t') = \delta\left(\xi - \xi' e^{(\ln t - \ln t')/2}\right) \theta(\ln t - \ln t'). \quad (44)$$

This implies that *the temporal translational invariance is restored in terms of the logarithmic time*.

Finally, let us examine the above-mentioned second and third results in view of the known properties of aftershocks. The second result is unsatisfactory since it has been reported in References [17,18] that processes of aftershocks are generally non-Markovian. Clearly, Markovianity of the present theory comes from that of Equation (3) and the hierarchical structure although subdynamics of a Markovian dynamics is not necessarily Markovian, in general. This point needs further investigations. On the other hand, the third result is intriguing since it shows how the subdynamics experiences *slowing down* in terms of the logarithmic time. This captures an element of *criticality*. Let us note that Green's function is actually a tri-variate function $G(\xi,t:\xi',t') \equiv g(\xi,\xi',t/t')$, as seen in Equation (44). Rewriting as $t \rightarrow t+t_w$ and $t' \rightarrow t_w$ with the waiting time $t_w$, this becomes

$$G(\xi,t+t_w:\xi',t_w) = g(\xi,\xi',t/t_w+1), \tag{45}$$

which depends on not only *t* but also the waiting time, showing nonstationarity of the evolution. The dependence on the waiting time here is specific: the larger the waiting time is, the slower the evolution in terms of the conventional time *t* becomes. That is, the subsystem exhibits the aging phenomenon (not in the two-time correlation function but in Green's function), implying that the subsystem has its own "internal clock". We mention that such a phenomenon has been discovered for the event-event correlation of real aftershocks [16,17]. It is also noted that the logarithmic time and the aging phenomenon appear in glassy dynamics [29].

## 4. Concluding Remarks

We have presented a theoretical approach to describing the Omori-Utsu law for earthquake aftershocks. Assuming fluctuating diffusivity effectively representing the system heterogeneity, we have examined the Fokker-Planck theory with the hierarchical structure, in which the load-state and diffusivity variables are the fast and slow degrees of freedom, respectively. In this way, we have extracted the information about the dynamics underlying the law that can be used in the stochastic process of aftershocks. Then, we have studied the evolution equation for the load state that is reduced from the Fokker-Planck equation. We have analyzed Green's function of that equation and have observed how the logarithmic time and the aging phenomenon naturally appear.

An additional point we make here is concerned with Equation (38) or, more fundamentally, Equation (37). As mentioned, these equations have the invariance under the individual rescaling transformations of the load-state variable and time. This symmetry makes the equations independent of the diffusivity (i.e. $D_0$) and leads to the deterministic transition between the load states. We have claimed that this symmetry has its origin in the scaling property of the conditional distribution in Equation (39). To see this somewhat in a wider context, let us look at, as an explicit example, the symmetric Lévy distribution indexed by $\alpha \in (0,2)$ [30]:

$$L_\alpha(\xi, t \mid D_*) = \frac{1}{2\pi} \int_{-\infty}^{\infty} dk \exp\left(-ik\xi - D_* t |k|^\alpha\right), \tag{46}$$

where $D_*$ stands for a generalized diffusion coefficient. The Gaussian case corresponds to the limit $\alpha \to 2^-$. This decays as a power law, $L_\alpha(\xi, t \mid D_*) \sim 1/|\xi|^{1+\alpha}$ and therefore its second moment is divergent. Accordingly, the diffusion property should be characterized not by the standard deviation but by e.g. the half width. We note that the conditional distribution in Equation (46) has the scaling property

$$L_\alpha(\xi, t \mid D_*) = \frac{1}{t^{1/\alpha}} \bar{L}_\alpha(\xi / t^{1/\alpha} \mid D_*) \qquad (47)$$

with the Lévy scaling function $\bar{L}_\alpha(x) = (2\pi)^{-1} \int_{-\infty}^{\infty} dk \exp\left(-ikx - D_* |k|^\alpha\right)$. This implies that the half width grows in time as $\sim t^{1/\alpha}$, exhibiting superdiffusion faster than normal diffusion $\sim \sqrt{t}$. Then, from Equation (47), it follows that

$$\frac{\partial L_\alpha(\xi, t \mid D_*)}{\partial t} = -\frac{1}{\alpha t}\left(\xi \frac{\partial}{\partial \xi} + 1\right) L_\alpha(\xi, t \mid D_*), \qquad (48)$$

which generalizes Equation (37). This equation still does not explicitly contain the (generalized) diffusion coefficient and therefore is invariant under the rescaling transformations of $\xi$ and $t$. It is however known that, in order to obtain the Lévy distribution as a solution of the Fokker-Planck equation, the operator $\partial^2 / \partial \xi^2$ should be fractionalized [30,31] and replaced by e.g. Riesz's fractional Laplacian. In general, not limited to the example in Equation (46), appearance of the deterministic transition is related to the scaling property of the conditional distribution.

*Note added*. After completing the present work, we have noticed Reference [32]. There, the authors discuss the Fokker-Planck theory with the hierarchical structure to the biological process experienced by the cell in connection with decision making. It shows how the theory can shed new light on its application to information theory.

**Author Contributions:** S.A. has formulated the problem and the all authors have equally contributed to the theoretical work. S.A. has organized the paper and the other two authors have agreed to publish it.

**Funding:** S.A. acknowledges the support from the Program of Fujian Province. The work of S.A. and D.A.T. has been supported by the Kazan Federal University Strategic Academic Leadership Program (PRIORITY-2030). N.S. is indebted to a Grant-in-Aid for Start-up Research of College of Science and Technology, Nihon University.

**Institutional Review Board Statement:** Not applicable.

**Informed Consent Statement:** Not applicable.

**Data Availability Statement:** Not applicable.

**Conflicts of Interest:** Not applicable.